\documentclass[aps,prl,twocolumn,showpacs,preprintnumbers,amsmath,amssymb,
superscriptaddress,nofootinbib]{revtex4}

\usepackage{graphicx} 
\usepackage{dcolumn}
\usepackage{bm}
\usepackage{epsfig}
\usepackage{epstopdf}
\usepackage{color}
\usepackage{ulem}
\DeclareGraphicsExtensions{.eps, .png, .jpg}

\begin{document}

\author{Flavio ~Giorgianni}
\affiliation{Department of Physics, University of Rome La Sapienza, P.le A. Moro 2, 00185, Rome, Italy}
\affiliation{Paul Scherrer Institute, SwissFEL, 5232 Villigen-PSI, Switzerland}
\author{Carlo ~Vicario} 
\affiliation{Paul Scherrer Institute, SwissFEL, 5232 Villigen-PSI, Switzerland}
\author{Mostafa ~Shalaby} 
\affiliation{Paul Scherrer Institute, SwissFEL, 5232 Villigen-PSI, Switzerland}
\author{Lorenzo Donato Tenuzzo} 
\affiliation{Department of Physics, University of Rome La Sapienza, P.le A. Moro 2, 00185, Rome, Italy}
\author{Augusto ~Marcelli} 
\affiliation{INFN-LNF, via E. Fermi 40, 00044 Frascati, Italy}
\affiliation{RICMASS, Rome International Center for Materials Science Superstripes, Via dei Sabelli 119A, 00185 Rome, Italy}
\affiliation{CNR - Istituto Struttura della Materia and Elettra-Sincrotrone Trieste, Basovizza Area Science Park 34149 Trieste, Italy} 
\author{Tengfei Zhang}
\affiliation{State Key Laboratory and Institute of Elemento-Organic Chemistry, Collaborative Innovation Center of Chemical Science and Engineering (Tianjin), Key Laboratory of Functional Polymer Materials and the Centre of Nanoscale Science and Technology, Institute of Polymer Chemistry, College of Chemistry, Nankai University, Tianjin, 300071, China}
\author{Kai Zhao}
\affiliation{State Key Laboratory and Institute of Elemento-Organic Chemistry, Collaborative Innovation Center of Chemical Science and Engineering (Tianjin), Key Laboratory of Functional Polymer Materials and the Centre of Nanoscale Science and Technology, Institute of Polymer Chemistry, College of Chemistry, Nankai University, Tianjin, 300071, China}
\author{Yongsheng Chen*}
\affiliation{State Key Laboratory and Institute of Elemento-Organic Chemistry, Collaborative Innovation Center of Chemical Science and Engineering (Tianjin), Key Laboratory of Functional Polymer Materials and the Centre of Nanoscale Science and Technology, Institute of Polymer Chemistry, College of Chemistry, Nankai University, Tianjin, 300071, China}
\author{Christoph ~Hauri} 
\affiliation{Paul Scherrer Institute, SwissFEL, 5232 Villigen-PSI, Switzerland}
\author{Stefano ~Lupi*}
\affiliation{Department of Physics, University of Rome La Sapienza, P.le A. Moro 2, 00185, Rome, Italy}

\title[Sample title]{Light and Sound in three-Dimensional Graphene Sponge}

\date{\today}

\begin{abstract}
\noindent Light modulation plays a key role in modern data transfer technologies providing many advantages, like low attenuation, large bandwidth and electric noise reduction.
In three-dimensional (3D) graphene sponge, we show that intensity modulated light can be transduced in acoustic waves through a highly efficient photo thermal-acoustic mechanism. As a first application of this effect, which is independent of light wavelength from infrared to ultraviolet, we demonstrate a photo-thermal based 3D graphene loudspeaker, permitting a full digital operation for frequencies from acoustic to ultrasound. The present results suggest a new pathway for light generation and control of sound and ultrasound signals potentially usable in a variety of new technological applications from high-fidelity loudspeaker and radiation detectors to medical devices.
\end{abstract}

\pacs{Valid PACS appear here}
\maketitle

\section{\label{sec:level1}Introduction}

Sound generation has been explored for millennia for communications, enjoyment and cultural reasons. Classical examples being drumheads and whistles for relatively long-distance exchange of information, and all kinds of musical instruments for religious and entertainement activities \cite{Sachs}. In contemporary society, sound generation and recording is even more important and efficient small-scale audio transduction systems like portable loudspeakers or wireless communication devices represent a cutting-edge technology for daily life.

For an efficient human audibility, an ideal speaker should generate a uniform sound pressure level (SPL) in the range 20 Hz-20 kHz. Most of loudspeakers commercially avalaible today, are based on a thin membrane connected with a voice coil nested in a permanent magnet. When alternating current passes through the coil, mechanical oscillations of the membrane are produced, leading adiabatic compressions and expansions of the surrounding air, $i.e.$ sound. However, the electro-mechanical resonances underlying the sound generation have an inherently narrow frequency response \cite{loudspeaker}. This represents the major limitation for high fidelity reproduction of a real sound.  
As a matter of fact, it is not possible to cover with a single mechanical loudspeaker a wide spectrum from acoustic region to ultrasound.\\

A first non-mechanical sound emission mechanism has been discovered in 1880 by A. G. Bell \cite{Bell1, Tyndall, Rontgen, Bell2}. 
Bell found that when a pulsed light beam shines both solids, liquids, and gases, an audible sound is generated.
His work generated a flurry of interest until recent years where photoacustic emission is used as a powerful spectroscopic technique in condensed matter physics \cite{Ball}.\\  
Another sound emission mechanism has been investigated by Arnold and Crandall \cite{Arnold}. Here, sound is produced through a conversion of heat energy in pressure waves and is called thermo-acoustic effect. More specifically, when an alternating current passes through a material a time dependent Joule heat is generated and then converted in sound which may extend beyond the acoustic region.

Although thermoacustic and photoacustic devices and their combination were conceived a long time ago, only recently the application of nanofabrication techniques allowed to produce acoustic devices based on ultrathin conductive structures \cite{CNT1,CNT2,goldnanowires,ito}, which permit to increase their conversion efficiency. However, the discovery of new, highly performant, photo-thermo materials could overcome the actual technical limits representing a cutting-edge technology for daily life. \\

Graphene, the two dimensional structure of carbon, reaches several records in term of mechanical strength \cite{Graphstrength}, electron mobility \cite{Graphelectronicproperties}, and thermal conductivity \cite{Graphthermalconductivity}. For these remarkable properties this material has been the subject of fundamental researches and it has used for new photonics and plasmonics devices \cite{graphene1,graphene2}.\\
Graphene is the ideal material for the photo thermo-acoustic conversion of light because it shows: i) A low heat capacity necessary to achieve large thermal gradients; ii) A high thermal conductivity to deliver rapidly heat to the surrounding gas; iii) An absorption coefficient, due to Dirac-particle-like electronic states \cite{Heinz}, practically indipendent of radiation wavelength from terahertz (THz) to ultraviolet (UV). Those unprecedented combination of optical, mechanical and thermal properties candidates graphene as one of the most useful material to transduce light in acoustic waves.\\ 

The first graphene thermo-acoustic device was based on a single layer graphene transferred on a substrate to achieve a good mechanical stability \cite{thermoacousticGraphene1}. In this case, however, the major part of heat relaxed from the absorbed light, is dissipated in the substrate thus reducing the conversion efficiency \cite{thermoacousticGraphene1}. This leakage effect can be reduced by using patterned substrate or nano-struttured graphene systems \cite{thermoacousticGraphene2,thermoacousticGraphene3}. Beside these constraints, single layer graphene based thermo-acoustic devices are able to operate in a wide frequency range up to tens of kHz.\\

\noindent Recently, it has been observed that the properties of 2-dimensional graphene can be extended on a macroscopic scale arranging individual graphene sheets in 3-dimensional (3D) monolithic structures \cite{reviewsponge}. In particular, graphene 3D sponges (G-sponge), while keeping the peculiar characteristics of single-layer graphene, add further interesting properties like an extremely lightness associated to a robust mechanical strength, and a highly repeatable compression and complete volume recovery in a wide temperature range both in air and in liquid without a substantial structure degradation \cite{gsponge, gspongepropulsion}.\\

In this paper, we report on the first light-driven loudspeaker based on photo thermo-acoustic effect in graphene-sponges. We found that a G-sponge sample is a highly efficient magnet-free, metal-free and contactless loudspeaker driven by light. In particular, our experiment reveals sound emission linearly dependent on light intensity and independent of light wavelength in a broad spectral range from infrared (IR) to UV. The graphene-sponge light-driven loudspeaker covers an acoustic range from 100 Hz to 20 kHz and, ultimately, it plays music from modulated light.\\

\section{\label{sec:level1}Results}
\noindent
\textbf{Acoustic waves in graphene sponge induced by modulated light.}\\
3D G-sponges were grown by an \textit{in situ} solvothermal process of graphene oxide (GO) sheets in ethanol (see methods). These sponges have an average transversal size of 1.5 cm and a thickness not less than 1 cm (see Fig. \ref{fig1}b). According to previous works \cite{gsponge,gspongepropulsion}, the 2D graphene electronic behavior due to a Dirac-like band structure is essentially preserved for the individual graphene sheets forming the G-sponge, while a weak band gap opens in proximity of the sheet interconnection regions \cite{gsponge,gspongepropulsion}. The G-sponge samples here investigated show a flat reflectivity from IR to UV of about 5 \%, and a density of 0.92 kg m$^{-3}$.\\
The scheme of the photo-acoustic setup is reported in Fig. \ref{fig1}a. Intensity modulated light was generated by LED sources powered by an electric signal properly amplified by a driver circuit. Emitted light enters in an anechoic chamber through an optical lens. This lens allows to focus the radiation onto the G-sponge sample with a spot size much smaller than the sponge surface (see Fig. 1b). The anechoic chamber was used both to acoustically isolate from environmental noise the experimental apparatous and to avoid internal sound reflection interference.\\
The emitted sound is finally acquired using a calibrated microphone (Samson-C02) placed behind the G-sponge at a fixed distance of 2.5 cm in the far-field region. \\
We investigated the sound emission as a function of excitation light wavelength covering the range from IR to UV by means of different nearly-monochromatic LEDs emitting at 1720 nm, 1050 nm, 780 nm, 280 nm, respectively, and through a broadband white LED emitting in the visible range from 400 nm to 750 nm. LED emission spectra are reported in the supplementary material.\\

We first discuss the results of acoustic generation in G-sponge by a sinusoidal modulated optical waveform. Fig. \ref{fig1}d shows the emitted sound driven by a sinusoidally modulated white LED at different modulation frequencies having a root-mean-square (rms) optical power of 20 mW. The excitation optical waveform is instead reported in Fig. \ref{fig1}a at 1 kHz. As one can observe, the acoustic wave is emitted at the modulating frequency of the optical excitation and its pressure amplitude increases with increasing frequency. This behavior is better shown in Fig. \ref{fig1}e, where the emitted sound pressure level (in dB), is reported by red dots as a function of the modulation frequency for an optical power of 20 mW. SPL monotonically increases versus the modulation frequency showing, for frequencies above 10 KHz, a saturation behavior.\\
Fig. \ref{fig1}f-g show the acoustic pressure \textit{vs.} the optical power $q_0$ at two different modulation frequency: 1 kHz and 10 kHz. As one can clearly see from those panels, the acoustic pressure linearly depends on the input power and it is practically independent of the excitation wavelengths. This is also true for a broadband white LED. All these data indicate the same sound generation efficiency for wavelengths from 1720 nm (IR) to 280 nm (UV).

The wavelength independent acoustic behavior previously observed is determined by two different factors: i) The photo-acoustic effect investigated in this paper is related to intercone transitions in the Dirac band structure of graphene, sketched in Fig. \ref{fig1}c. This results in a photon wavelength independent absorption coefficient \cite{Heinz}; ii) At these photon wavelengths the dominant electronic cooling channel is the non-radiative electron-phonon scattering \cite{gspongepropulsion}. This implies, approximately, that the whole optical energy is converted into Joule heating.

\noindent For higher photon energies, however, other non-radiative mechanisms take place, such as the emission of electrons by Auger-like effect \cite{gspongepropulsion} that can affect the efficiency of photo-thermal conversion.\\
The increase of lattice temperature due to the electron relaxation towards phonons leads to a thermal wave inside the G-sponge. Due to the low heat capacity of graphene this thermal energy is rapidly dissipated towards the air layers adjacent to the graphene interfaces. This generates a quick expansion of the air layers which act as an acoustic piston on the rest of surrounding gas column producing a pressure wave, $i.e.$ a sound wave.\\
As a result, light driven acoustic emission mechanism in G-sponge can be described in terms of a combination of two distinct processes characterized by different timescales: A photo-thermal and a thermo-acoustic process. The energy relaxation in the photo-thermal mechanism is driven mainly by the electron-phonon scattering and occurs on a fast time scale of tens of picoseconds \cite{phononscattering1,phononscattering2}. Through this effect, light energy absorbed by electronic intercone transitions, is mainly transduced in an out-of-equilibrium phonon populations.
The thermal energy here accumulated is then transfered over a microsecond scale to the surroinding air. 
Assuming that the light energy absorbed is completely converted in thermal heating (see supplementary material for the thermal properties of the G-sponge under illumination), the acoustic generation can be described through a model proposed by Hu $et$ $al.$ \cite{thermoacousticmodel}, which calculates the thermo-acoustic emission from a solid. 

\noindent The effective thermal length $\lambda_G=\sqrt{\alpha_G/(\pi F)}$, where $\alpha_G=2.25\cdot10^{-6} m^2s^{-1}$ is the graphene-sponge thermal diffusivity \cite{thermophone} and $F$ is the acoustic frequency, represents the maximum spatial extension of the thermal waves in the G-sponge.
For the 3D G-sponge investigated in this paper and for acoustic frequency higher than a few of Hz, $\lambda_G$ is less than the sample thickness.
In this limit, the sound pressure (rms value), according with the model of Hu $et$ $al.$, can then be expressed as \cite{thermoacousticmodel}:

\begin{equation}
p=\frac{R_0}{r_0}\frac{\gamma-1}{v_g}\frac{e_g}{Me_g+e_{G}}(1-R)I_0,
\label{eq1}
\end{equation}

\noindent where $e_g$ and $\gamma$ are the thermal effusivity ($i.e.$  the rate at which a specific material can exchange heat with the surrounding environment), and the heat capacity ratio of the surrounding gas, respectively. 
M is a frequency dependent factor which approaches 1 at high frequency. $v_g$ is the sound velocity in the gas, and $e_{G}=\sqrt{\kappa \rho_{G} C_G}$ is the thermal effusivity of the G-sponge. Here, $\kappa=\alpha_G\rho_{G}C_G$ is the sponge thermal conductivity, $\rho_{G}$ its mass density and $C_G=690$ Jkg$^{-1}$K$^{-1}$ its heat capacity whose value has been measured in Ref. \cite{thermophone}. $R\sim0.05$, is the G-sponge reflectivity, taking into account reflection loss. $I_0=q_0/A$ is the input light intensity where $q_0$ the optical input power and A the illuminated sponge  surface.\\
In Eq.1, R$_0$/r$_0$, where R$_0=FA/v_g$ is the Rayleigh distance, and $r_0$ is the distance between the sound source and the microphone, provides the right far-field limit where the experiments have been performed.\\
As shown in Fig. \ref{fig1}e by a blue dashed line, Eq. \ref{eq1} quantitatively reproduces the sound pressure as a function of frequency without the use of free parameters. Let us notice, that the overestimate at high pressure of theory with respect to experimental data, can be associated mainly to a reduced efficiency of the microphone.\\ 
Moreover, the calculated acoustic pressure has a linearly dependence on the optical power $q_0$ as experimentally observed. The slope calculated by Eq. \ref{eq1} is 0.07 (0.75) Pa/W at 1 kHz (10 kHz) which is very close to 0.08 (0.81) Pa/W experimentally obtained (see Fig. \ref{fig1}f and g). The small difference between data and calculation are related to the incertitudes in the G-sponge thermal parameters entering in Eq. 1.\\

\noindent
\textbf{Acoustic waves emission from compressed G-sponge}\\
A crucial parameter of the photo thermo-acoustic effect is the thermal effusivity $e$. Indeed, as observed in Eq. \ref{eq1}, in order to obtain an efficient photo thermo-acoustic effect the effusivity of a material must be comparable to that of the surrounding gas. The effusivity in conventional bulk conductors is very high thus they are not suitable for thermo-acoustic applications. This limitation has been partially overcame by fabricating thin and nanostructured materials in which the mass density and the heat capacity $per$ unit area (HCPUA) are strongly reduced \cite{goldnanowires,ito,metalwire,silvernanowires}.\\

\noindent In order to investigate the dependence of the acoustic emission on the G-sponge morphology, we have applied a huge compressive strain to the G-sponge sample. SEM data of an uncompressed and compressed G-sponge over 99\% of applied strain are represented in Fig. \ref{figura2}. The as-grown sponge consists of a frame of interconnected micro-scale voids, resulting in an open-cell structure. With an applied compressive strain larger than 99\%, the G-sponge enters in a plastic deformation regime, where the void space shrinks and the sample becomes denser as shown in Fig. \ref{figura2}b. Notably, the graphene-microvoid walls originally randomly distributed, stack in nearly parallel-aligned arrays perpendicular to the compressive strain direction.
Ultimately, a 99\% compression, results in an increased mass density of nearly a factor 10 in comparison with the uncompressed sample.

\noindent Fig. \ref{figura2}c shows the sound pressure level (blue points), versus frequency, generated from a compressed G-sponge illuminated by a sinusoidally modulated white LED at 20 mW.
SPL, as for the uncompressed G-sponge sample, increases with frequency. However, the compressed G-sponge shows lower sound pressure (about a factor 2, blue points) than the uncompressed one (black line), for the same input optical power. This means a reduction in the photoacoustic conversion efficiency of about a factor 4. A reduced efficiency is also observed in Fig. \ref{figura2}d,e which show the acoustic pressure \textit{vs} the input optical power at two different frequencies (1 kHz and 10 kHz), compared with the uncompressed results (colored points and continous blue line, respectively).\\

At a very high compressive strain, anisotropy is induced into the sponge, and a growing alignment of the graphene sheets perpendicular to the direction of compression occurs (see Fig. \ref{figura2}b). This leads to a final plastic deformation of the sample determining a long-range (millimeter) order, keeping intact, instead, the short spatial structure (micrometer) of the graphene sponge network, as confirmed by SEM image in the insets in Fig. \ref{figura2}a and Fig. \ref{figura2}b. In particular, the macroscopic order does not affect the thermal and optical properties of the individual sheets. Therefore, the reduction of the sound emission efficiency is mainly due to the increase of the G-sponge density which reduces the sponge effusivity $e_{G}$.
This behavior is further confirmed by putting the density of compressed G-sponge in Eq. \ref{eq1}. The numerical results, shown through a red curve in Fig. \ref{figura2}c, well reproduce the experimental data.\\

\noindent
\textbf{Coherent acoustic wave emission}\\
As previously described, a sound wave is generated when graphene-sponge samples are shined by a modulated light. The sound power is linearly dependent on the light intensity and independent of the light wavelength from infrared to ultraviolet. 

In order to investigate the time scale formation of the sound wave, in this subsection we study the acoustic generation as a function of the rise time of a square-wave optical-excitation. The square wave from a white LED at different rise times are reported in Fig. \ref{figura3}a by red lines, while the corresponding sound waves are shown by colored lines. A narrow acoustic pulse appears when the optical rise time is less than a few tens of $\mu$s. For these time-scales a broadband pulse is generated in the frequency domain, covering the whole acoustic region, as shown in Fig. \ref{figura3}b. An increase of the rise time determines a temporal broadening of the acoustic pulses, which reflects in a reduction of their intensity and spectral distribution in the frequency domain.\\
The acoustic wave emitted, in the same optical conditions, for a rise time less than 1 $\mu$s and a repetition rate frequency of 100 Hz, is shown in Fig. \ref{figura3}c as a function of time. Each period has two optical edges, corresponding to the rising and falling of the light wavefront, which are labeled in Fig. \ref{figura3}c, by a sign (+) and (-), respectively. An expansion pressure wave is then generated in correspondence of the optical rise time. This is due to a fast heating of the graphene sponge above room temperature (the thermal properties of graphene sponge under illumination have been reported  in SI). At the optical falling time, a fast cooling process towards room temperature is instead produced inducing a compression pressure wave (Fig. \ref{figura3}c). The expansion and compression waves actually present an exact specular shape.\\
Increasing the repetition rate of the optical square wave for a fixed optical power, the time dependence of the acoustic waveform changes significantly (see Fig. \ref{figura3}d). This effect can be ascribed to a progressive temporal superposition of the expansion and compression waves. In particular, the heating-cooling waves are constructively overlapped when the frequency approaches 10 kHz resulting in an effective increase of the acoustic pressure. For higher frequencies instead, heating-cooling waves begin to distructively interact decreasing the emitted acoustic pressure as observed, for instance, at 15 kHz in Fig. \ref{figura3}d.\\ 

In order to simulate this overlapping effect we performed a numerical calculation. In this simulation (see Fig. 3e), the acoustic pulses at 100 Hz (see Fig. \ref{figura3}c), which can be considered as non-interacting, have been superimposed with alternating signs. By increasing the modulation frequency up to 15 KHz $i.e.$ by reducing the period among two consecutive pulses, the simulation accurately reproduces both the shapes and the intensities of the measured acoustic waves. This superposition effect clearly influences the sound spectral distribution as observed in Fig. \ref{figura3}f. From a broadband spectrum generated by a single acoustic pulse (at 100 Hz), the sound spectral distribution evolves in a comb spectrum of odd harmonics of the modulation frequency (from 1 kHz to 15 kHz). This agrees to the fact that the Fourier expansion of a square wave contains only odd harmonics.\\

\noindent
\textbf{Light-driven loudspeaker}\\
The first light-driven loudspeaker was proposed in 1983 by W. F. Rush \textit{et al} which, following the Bell's idea, generated sound through an intensity modulated laser beam in a gas \cite{photospeaker}. This device was able to overcome the mechanical limitations of the conventional loudspeaker achieving high fidelity sound reproduction. However, due to the very weak gas photoacoustic efficiency and therefore the need of a powerful laser system, the spread of this device was limited.The light-sound transducer based on G-sponges investigated in this manuscript has instead, due to its high efficiency and fidelity, a huge potentiality for new applications, in particular as loudspeaker and broadband light detector.\\
As discussed above, LED sources are very fast but their current-voltage response shows a weak nonlinear behavior which leads to a small distortion of the light output waveform. This represents a drawback in the generation of acoustic waves using the light intensity modulation scheme. As a result, one generates both the fundamental modulation frequency and high order harmonics which may distort consequently the acoustic output (see Fig. \ref{figura4}a).\\
In order to generate a pure tone and achieving a high fidelity in sound reproduction, G-sponge loudspeaker can be driven by a light pulse density modulation (PDM) method \cite{tin}. In PDM mode, an analog input signal is encoded in sequential pulse trains with constant height and a relative temporal pulse density that corresponds to the analog signal's amplitude.
A sinusoidal analog waveform (5 kHz), is encoded by PDM in a digital signal which drives the LED (Fig. \ref{figura4}b). The generated light pulse trains measured by a photodiode, where a single light pulse has a FWHM of 1 $\mu$s, are shown in Fig. \ref{figura4}c. The acoustic output signal is instead shown in Fig. \ref{figura4}d. We can see that from a light pulse density modulated signal the original analog signal is well reproduced.\\
The generated acoustic wave in frequency domain is reported in Fig. \ref{figura4}e. The first harmonic intensity, which, in the intensity modulation scheme, is nearly 10 \% of the fundamental one, decreases to about 1 \% in the PDM mode.  Moreover, the distortion in the light driven loudspeaker based on G-sponge is lower than in a commercial moving-coil loudspeaker as shown in  Fig. \ref{figura4}e. Even better results can be obtained by reducing the duration of the single light pulse and increasing the digital sampling of the analog input waveform.\\
Fig. \ref{figura4}f, g show the time-frequency plots for linear swept-frequency sine input signal from 200 Hz to 20 kHz by using light intensity modulation and PDM mode, respectively. Let us further observe that high order harmonics are strongly suppressed using the PDM method.\\
Therefore, G-sponge loudspeaker is absolutely compatible with the PDM sampling technique, permitting a full digital operation integrable with other electronic devices and providing an ultra-high sound fidelity with frequencies from acoustic to ultrasound. The capability of a G-sponge light speaker in reproducing a real audio file is shown in Supplementary Video 1 and 2. Here, G-sponge based-speaker is able to play songs by intensity modulated light with a 20 mW rms power generated from a commercial with LED.

\noindent In this work we have presented an unprecedented concept of photo-acoustic speaker taking advantage of the thermodynamic, mechanical and optical properties of a free-standing three-dimensional graphene sponge structure. 
The use of graphene sponges, which have a  density similar to air, provides an unparalleled lightness and mechanical stability of the loudspeaker while the absence of the substrate prevents thermal leakage.
The light-driven acoustic generation allows to emit sound from far distances without any type of physical connection. Moreover, the heat distribution on the sponges can be fully controlled by the spatial distribution of the light giving rise to a scalable device with an acoustic power linearly dependent on the illuminated surface.
The present results suggest that the light-driven acoustic generation in graphene sponge could be widely used in a variety of new technological applications from sound and ultrasound production, radiation detectors and medical devices.

\section{Author Contributions} 
Tengfei Zhang, Kai Zhao, Yongsheng Chen fabricated and characterized 3D graphene sponges. F. Giorgianni, S. Lupi, A. Marcelli, M. Shalaby, L. D. Tenuzzo, C. Vicario, carried out the infrared, visible and ultraviolet experiments. Data analysis has been performed by F. Giorgianni and L. D. Tenuzzo. F. Giorgianni and S. Lupi planned and managed the project with inputs from all the co-authors. F. Giorgianni and S. Lupi wrote the manuscript. All authors extensively discussed the results.

\section{Additional Information}
The authors declare no competing financial interests. 
Correspondence and requests for materials should be addressed to S.L. (stefano.lupi@roma1.infn.it) and Yongsheng Chen (yschen99@nankai.edu.cn).

\bibliography{aipsamp}

\begin{widetext}

\begin{figure*}
  \centering \includegraphics[scale=.55]{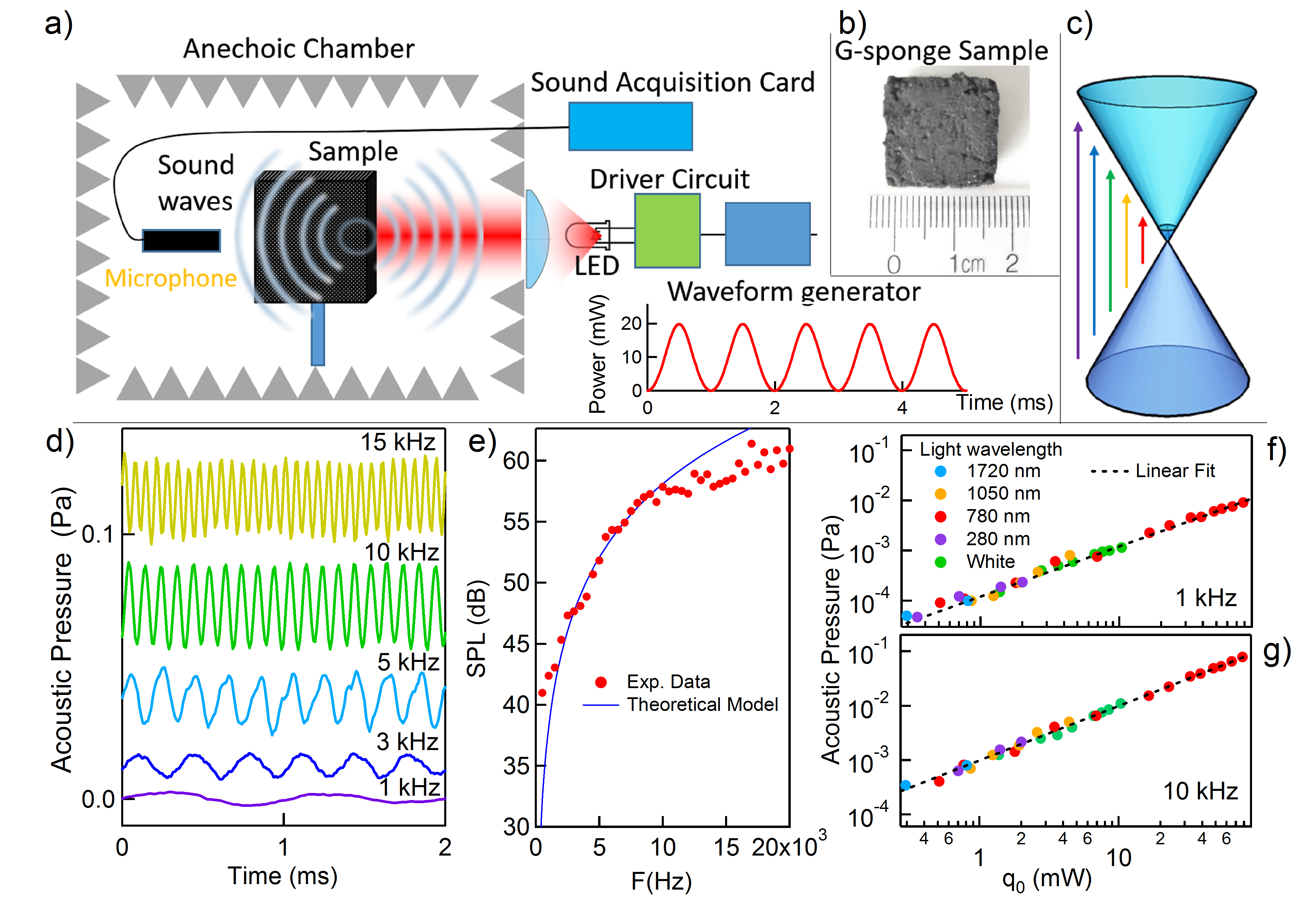}
  \caption{\textbf{Sound generation from Graphene Sponge} a) Scheme of experimental setup for acoustic generation measurements: Intensity modulated light at different wavelengths and modulation frequencies is emitted by a LED powered by a driver circuit coupled with a waveform generator. The light then enters through a lens inside the anechoic chamber irradiating the G-sponge. b) View of the G-sponge sound emitter. c) Linear (Dirac) electronic band dispersion in graphene: The arrows represent the inter-band transitions under an optical excitation. The electrons involved relax by electron-phonon interaction inducing a lattice heating on fast (ps) time-scale. The fast-rise thermal gradient in the G-sponge network generates acoustic waves. d) Acoustic emission in time domain at different frequencies of a sinusoidally modulated white light with a rms power of 20 mW (vertical offset added). e) The generated sound pressure level as a function frequency with at 20 mW input power of sinusoidal light. Red dots represent the experimental data and the blue solid line corresponds to a model of thermo-acoustic emission, which is described in the main text. f)-g) Generated acoustic pressure vs rms optical power at 1 kHz and 10 kHz, respectively. Dots with different colors indicate different wavelengths of excitation light: 1710 nm, 1050 nm, 780 nm, 280 nm, and white light. Dashed line is a linear fit. The clear linear dependence of the acoustic pressure $vs.$ the optical power is independent of the light wavelength across the visible range.}
\label{fig1}
\end{figure*}

\begin{figure*}
 \centering \includegraphics[scale=.9]{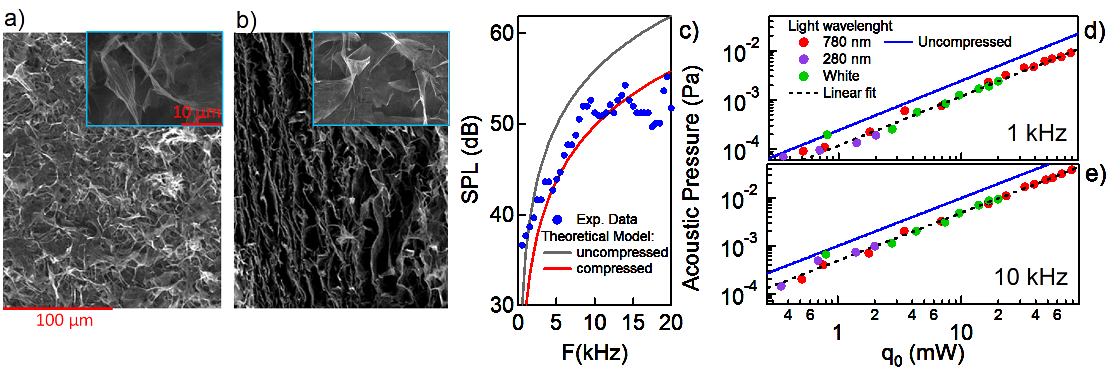}
  \caption{\textbf{Effect of compressive strain on the sound generation efficiency in G-sponge.} SEM images: a) G-sponge surface b) G-sponge surface parallel to the compressive direction for a strain over 99 \% (plastic regime). Insets in a) and b) show magnified SEM images for G-sponge and compressed G-sponge, respectively. At short-range the network morphology and the physical parameters of graphene sheets remain approximatively unchanged. c) Generated SPL from compressed G-sponge by a sinusoidal optical waveform with 20 mW of rms power: Blue dots are experimental data, red solid line results from the thermo-acoustic model for compressed G-sponge (see main text), while red solid line corresponds to the thermo-acoustic emission for the uncompressed G-sponge. d)-e) Generated acoustic pressure $vs.$ the rms optical power for the compressed G-sponge at 1 kHz and 10 kHz, respectively. Dots with different colors indicate different wavelengths of excitation light: 780 nm, 280 nm, white line. Dashed line is a linear fit. As a result, the conversion efficiency decreases with respect to uncompressed G-sponge (blue line).}
 \label{figura2}
\end{figure*}

\begin{figure*}
  \centering \includegraphics[scale=0.75]{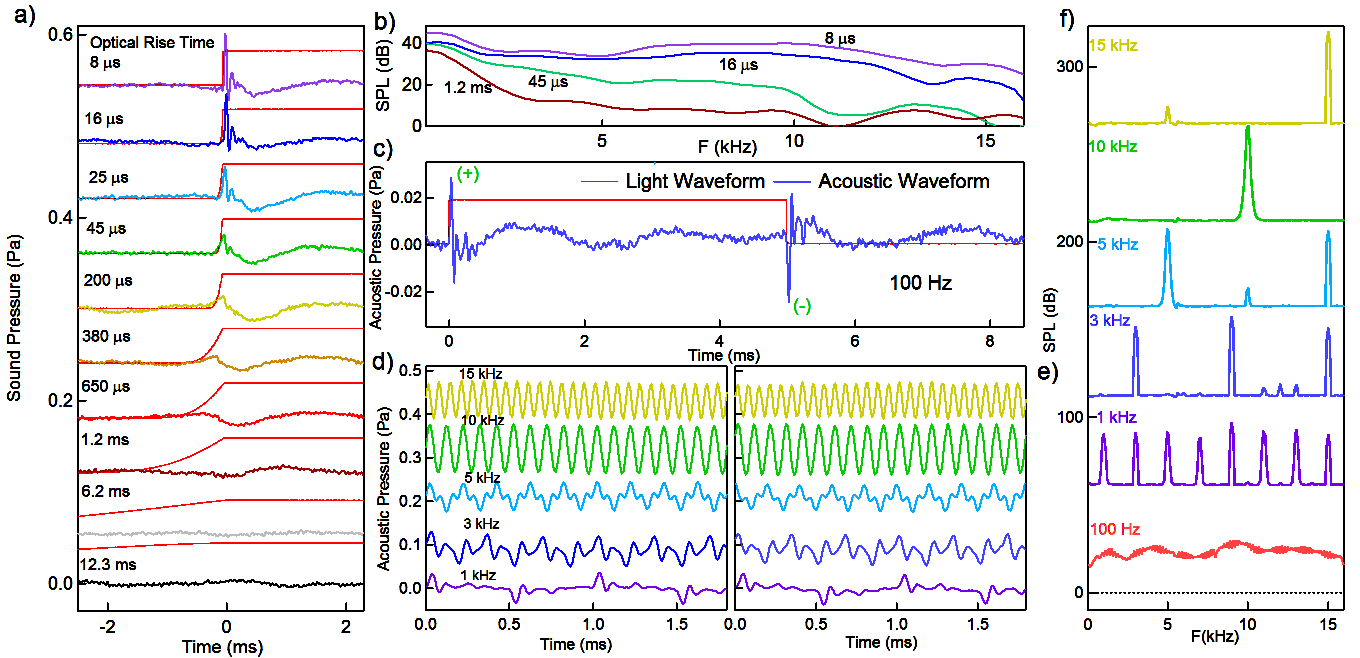}
   \caption{\textbf{Broadband acoustic generation from optical square-wave in G-sponge: time and frequency domain.} a) Temporal profile of the acoustic pulse generate in G-sponge as a function of rise time of an optical square-wave which is shown by a red curve (vertical offset added). b) Related acoustic spectra for different rise time. c) Generated acoustic wave (blue curve) by an optical square-wave (red curve) with 100 Hz modulating frequency. d) Time domain signal of G-sponge sound at different modulation frequencies of the optical square wave. A vertical offset has been applied to the curves for a clearer view. e) Numerical simulation of the temporal overlapped acoustic waves. f) Evolution of the emission spectrum $vs.$ the amodulation frequency. All measurements were performed with a white LED optical waveform with a rms power of 20 mW.}
\label{figura3}
\end{figure*}

\begin{figure*}
 \centering \includegraphics[scale=0.5]{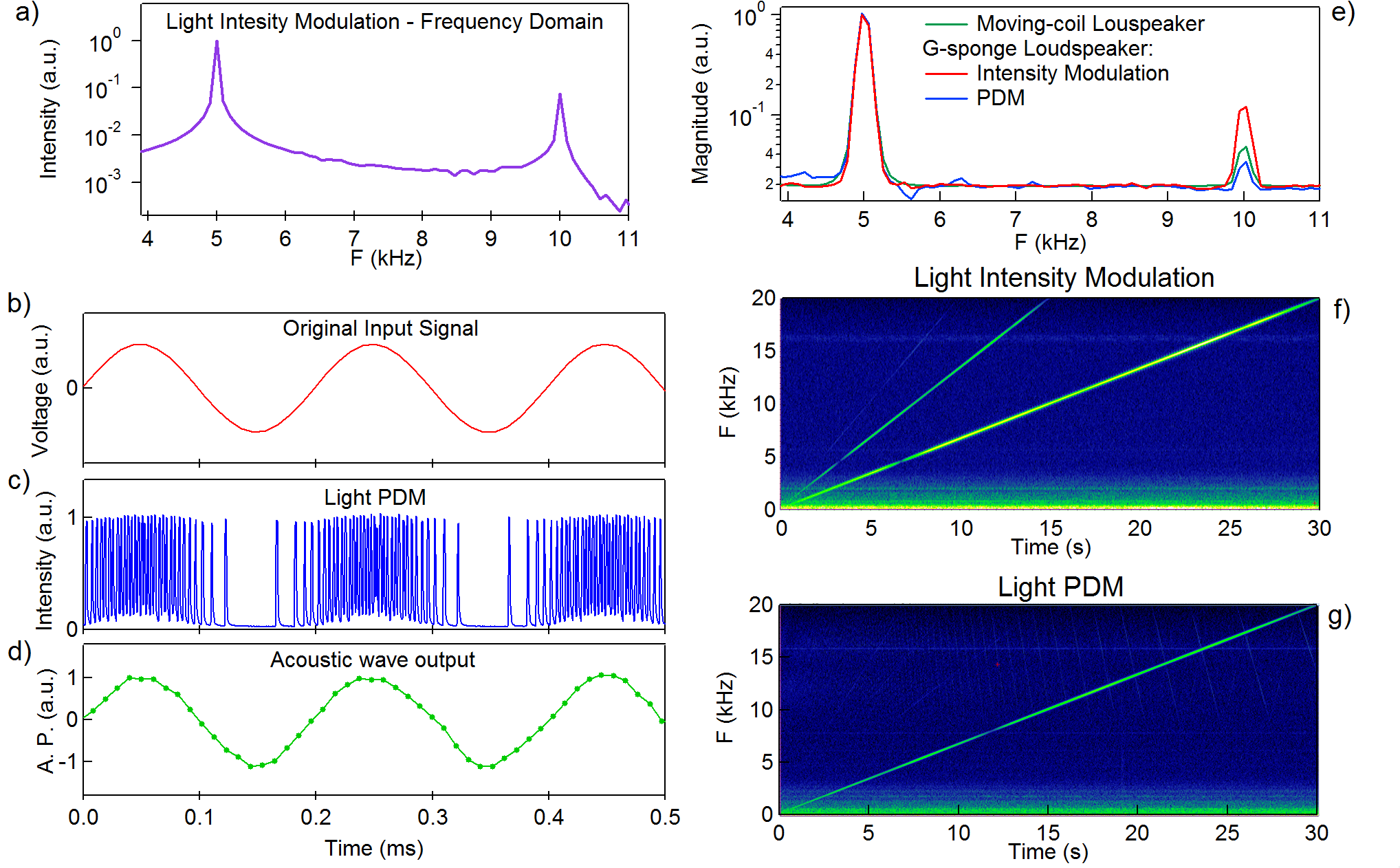}
\caption{{\bf Light Pulse Density Modulation:} a) Sinusoidally intensity modulated light in frequency domain emitted from a white LED. Modulation frequency is 5 kHz. The LED response induces a distortion of emitted light waveform and consequently a first harmonic is observed at 10 KHz.  b) Waveform of an analog input signal (5 kHz). c) The corresponding light pulse trains from PDM-digitized analog signal. d) Measured acoustic output signal. This full digital operation mode very well reproduces the analog input signal. e) Acoustic wave emitted in the G-sponge through the intensity modulation mode (red curve), PDM mode (blue curve). The corresponding acoustic wave emitted from a commercial moving-coil loudspeaker at 5 kHz (green curve). f) Time-frequency plot of generated acoustic waves of linear swept-frequency input signal by intensity modulation mode (see main manuscript) g) Corresponding generation acoustic waves by PDM mode. As one can observe the harmonic distortion is strongly suppressed.}
\label{figura4}
\end{figure*}

\end{widetext}

\end{document}